\documentclass[9pt,twocolumn,twoside]{osajnl}

\journal{ol} 

\setboolean{shortarticle}{true}

\title{Anomalous multi-ramp fractional vortex beams with arbitrary topological charge jumps}

\author[1,2,3]{Jun Zeng}
\author[2]{Zhiheng Xu}
\author[3,4]{Chengliang Zhao}
\author[2,3,5]{Yangjian Cai}
\author[1,6]{Greg Gbur}

\affil[1]{Department of Physics and Optical Science, The University of North Carolina at Charlotte, Charlotte, North Carolina 28223, USA}
\affil[2]{Shandong Provincial Engineering and Technical Center of Light Manipulations \& Shandong Provincial Key Laboratory of Optics and Photonic Device, School of Physics and Electronics, Shandong Normal University, Jinan 250014, China}
\affil[3]{School of Physical Science and Technology, Soochow University, Suzhou 215006, China}

\affil[4]{E-mail: zhaochengliang@suda.edu.cn}
\affil[5]{E-mail: yangjiancai@suda.edu.cn}
\affil[6]{E-mail: gjgbur@uncc.edu}

\begin{abstract}
Traditional fractional vortex beams are well-known ``jump'' beams: that is, their net topological charge jumps by unity as the effective topological charge of the source passes a half-integer value. Here, we propose an anomalous multi-ramp fractional vortex (AMRFV) beam. Unlike the traditional fractional vortex beams, an AMRFV beam can be designed to have arbitrary jumps in topological charge at any critical threshold of the source charge.  We walk through some examples of  AMRFV beams using simulations and present a clear interpretation of the multi-jump characteristic based on the evolution of phase singularities.
\end{abstract}

\setboolean{displaycopyright}{true}

\begin{document}

\maketitle

~\\
The study and application of beams with helical wavefronts and the associated singularity of phase at their core has become an important aspect of the field now known as \emph{singular optics} \cite{gjg:so:2017}.  In 1992, such vortex beams with an azimuthal order $l$ were shown by Allen \emph{et al.} \cite{lamwbrjcsjpw:pra:1992} to possess an orbital angular momentum (OAM) of $l\hbar$ per photon. In fact, the phase in a counterclockwise path around every vortex in a general wavefield changes by $2\pi l$, where $l$ is referred to as the topological charge (TC) of the vortex and always takes on an integer value. The discrete nature of topological charge has generated much interest in its use as an information carrier in optical communications \cite{ggrkt:josaa:2008}, among other applications.

Though the topological charge can only take on integer values, it is possible to formally generate a ``fractional vortex beam'' by the use of a fractional spiral phase plate that induces only a fraction of a $2\pi$ phase circulation around the beam axis. This was studied in detail by Berry in 2004 \cite{mvb:joa:2004} (though an early look at such possibilities appeared in 1995 \cite{ivbmssmvv:oc:1995}), who also showed how the topological charge of a modified plane wave ``jumps'' by unity when the effective charge of the fractional plate takes on a half-integer value . This result was demonstrated experimentally soon after \cite{jleymjp:njp:2004}. Since Berry's pioneering work, considerable attention has been paid to such fractional vortex beams, both coherent and partially coherent cases \cite{cgyyzh:oc:2010,yfqlxwwzlc:pra:2017,jzxlfwczyc:oe:2018,slbsxzzbwg:oe:2018,jwlwxyjzsz:oe:2019,jzclhwfwczggyc:oe:2020}. 

The early work on fractional beams imagined creating them with a conventional spiral phase plate \cite{mbrcmkjw:oc:1994}, with a single adjustable spiral ramp in the azimuthal direction providing the fractional phase twist and an effective fractional topological charge at the source. But in a demonstration that fractional vortex beams create new topological charge through a ``Hilbert Hotel'' mechanism, Gbur \cite{gg:o:2016} showed that an adjustable multi-ramp spiral phase plate, with $m$ ramps in the azimuthal direction instead of one, could create a jump of $m$ in topological charge as the source charge is increased, corresponding to $m$ rooms being simultaneously freed in Hilbert's Hotel. This hinted at the possibility that even more sophisticated fractional spiral phase plates could be designed which generate an arbitrary jump of topological charge with an increase of the fractional source charge. It is reasonable to believe that the future requirements for vortex beams with TC structure will develop towards diversity and richness, especially for particle trapping \cite{cgyyzh:oc:2010}, information security \cite{xfhrmg:np:2020}, and optical communication \cite{yrzwplllgxhhzzyynaawmpjlnasarbmtibdmanaew:ol:2016}. Therefore, the generation and modulation of vortex beams with complex and diverse TC structures will be a topic of constant attention.

In this Letter, we modify the transmission function of the spiral phase plates from Ref.~\cite{gg:o:2016} to design an anomalous multi-ramp spiral phase plate (AMRSPP) which generates an anomalous multi-ramp fractional vortex (AMRFV) beam, and demonstrate its rich and varied TC jump characteristics with simulations.  We show that it is possible to design a plate with a single tunable parameter -- the effective source charge -- that will have topological charge jumps of arbitrary size that appear in any order desired.

\begin{figure}[htbp]
\setlength{\abovecaptionskip}{7pt}
\setlength{\belowcaptionskip}{7pt}
\centering
   \includegraphics[width=\linewidth]{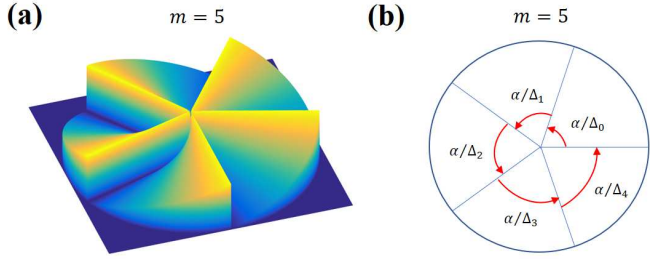}
   \caption{(a) Illustration of an AMRSPP with $m=5$, (b) the top view of Fig.1. (a) with detailed structural distribution.}
\label{amrspp:fig}
\end{figure}

Figure \ref{amrspp:fig} shows a designed AMRSPP involving three types of adjustable parameters, including the number of multi-ramp sections $m$, the input fractional topological charge $\alpha$, and the fractional phase control coefficients ${\Delta_p}\left( {p = 0,1,2,...,m - 1} \right)$ in each ramp section. The transmission function of an AMRSPP is then given by
\begin{equation}  
T\left( \theta  \right) = \exp \left[ {\frac{{i\alpha \left( {\theta  - 2\pi p/m} \right)}}{{{\Delta_p}}}} \right], \frac{{2\pi p}}{m} \le \theta  < \frac{{2\pi \left( {p + 1} \right)}}{m},
\label{plate:def}
\end{equation}
where $\theta$ denotes the azimuthal angle. 

In analogy with a conventional spiral phase plate, we envision the AMRSPP as a transparent plate with a polymer replicated on a glass substrate, and the thickness of each ramp section varies azimuthally. Here, the phase increases by $2\pi \alpha /{m\Delta_p}$ in each ramp section, which is quite different from the simple design in Ref.  \cite{mbrcmkjw:oc:1994} and that possessing the same phase change rate in each ramp section in Ref. \cite{gg:o:2016}. Using such an AMRSPP, we can attach a specific phase twist to the beam and generate an AMRFV beam with complex and diverse TC structures. When $m=1$ and ${\Delta_0}{\rm{ = }}1$, the model designed by Eq.~(\ref{plate:def}) reduces to a conventional spiral phase plate, which is used to generate a traditional fractional vortex beam. For $m \ge 2$ and all ${\Delta _p}{\rm{ = }} 1$, the model designed by Eq.~(\ref{plate:def}) reduces to a multi-ramp spiral phase plate, which is used to realize a fractional vortex beam with multi-unit (i.e., $m$-unit) TC jump \cite{gg:o:2016}.

It is to be noted that we can generalize our spiral phase plate even further: the ramp sections can be taken to have different azimuthal widths, and the height of each ramp section can be taken to vary independently.  Here, we primarily consider plates with a single variable parameter -- the ``source charge'' -- and demonstrate that we can create Hilbert Hotel style jumps of topological charge in any order and any degree even under this restricted case. 

For mathematical convenience, we consider the illumination of such a phase plate by a normally incident plane wave. According to the Fresnel diffraction principle and the Collins formula \cite{swdz:mo:2000}, the field of a monochromatic plane wave with unit amplitude passing through a spiral phase plate of integer topological charge $n$ [designed to have a transmission function ${T_0}\left( \theta  \right){\rm{ = }}\exp \left( {in\theta } \right)$ ] is expressed as
\begin{equation}  
\begin{split}
{E_n}\left( {\rho ,\theta ,z} \right) =& \sqrt {\frac{{\pi {\rho ^2}}}{2}} \exp \left( {ikz} \right)\exp \left( {in\theta } \right)\exp\left( {i{\rho ^2}} \right){( - i)^{\left| n \right|/2}}\\
&\times \left[ {{J_{\frac{{\left| n \right| - 1}}{2}}}\left( {{\rho ^2}} \right) - i{J_{\frac{{\left| n \right| + 1}}{2}}}\left( {{\rho ^2}} \right)} \right],
\end{split}
\label{integervortex}
\end{equation}
with
\begin{equation}  
\rho {\rm{ = }}\sqrt {{\xi ^2}{\rm{ + }}{\eta ^2}} ,{\rm{ }}\xi {\rm{ = }}\sqrt {k/4z} x,{\rm{ }}\eta {\rm{ = }}\sqrt {k/4z} y,
\end{equation}
where $z$ and ${\boldsymbol{r}} {\rm{ = }}\left( {x,y} \right)$ denote the distance from the phase plate and the transverse position vector in a plane of $z$, respectively. $k$ is wavenumber and $\rho$ is the scaled dimensionless variable. ${J_n} \left(\cdot\right)$ denotes the $n$th order Bessel function of the first kind.  

Following Berry \cite{mvb:joa:2004} and Gbur \cite{gg:o:2016}, the transmission function of AMRSPP can be expanded into the Fourier series of integer order
\begin{equation}  
T\left( \theta  \right) = \sum\limits_{n =  - \infty }^\infty  {{C_n}\exp \left( {in\theta } \right)} ,
\end{equation}
with
\begin{equation}  
{C_n} = \frac{1}{{2\pi }}\sum\limits_{p = 0}^{m - 1} {\frac{{\exp \left( { - i2\pi pn/m} \right)}}{{i\left( {\alpha /{\Delta _p} - n} \right)}}\left\{ {\exp \left[ {i\frac{{2\pi }}{m}\left( {\frac{\alpha }{{{\Delta _p}}} - n} \right)} \right] - 1} \right\}} .
\end{equation}

Thus, by applying Eq.~(\ref{integervortex}), the field of an AMRFV beam can be obtained as follows
\begin{equation}  
{E_\alpha }\left( {\rho ,\theta ,z} \right) = \sum\limits_{n =  - \infty }^\infty  {{C_n}{E_n}\left( {\rho ,\theta ,z} \right)} .
\label{amrfv:field}
\end{equation}

To study the topological behavior of AMRFV beams, we consider the net topological charge (also called total vortex strength), defined as \cite{mvb:joa:2004}
\begin{equation}  
\begin{split}
t &= \mathop {\lim }\limits_{\rho  \to \rho_0 } \frac{1}{{2\pi }}\int_0^{2\pi } {d\theta  \cdot \frac{\partial }{{\partial \theta }}} \arg \left[ {{E_\alpha }\left( {\rho ,\theta ,z} \right)} \right]\\
 &=\mathop {\lim }\limits_{\rho  \to \rho_0 } \frac{1}{{2\pi }}\int_0^{2\pi } {d\theta  \cdot {\mathop{\rm Re}\nolimits} \left\{ { - i \cdot \frac{{\partial {E_\alpha }\left( {\rho ,\theta ,z} \right)/\partial \theta }}{{{E_\alpha }\left( {\rho ,\theta ,z} \right)}}} \right\}.}
\label{ntc}
\end{split}
\end{equation}

\begin{figure}[ht] 
\centering
\includegraphics[width=\linewidth]{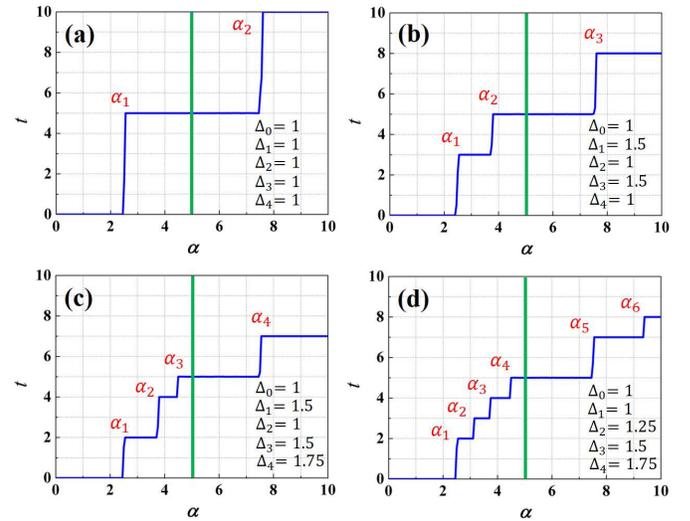}
\caption{Net topological charge $t$ of an AMRFV beam as a function of $\alpha$ with different fractional phase control coefficients $\Delta_p$ for $m=5$, calculated by numerically evaluating the integral of Eq.~(\ref{ntc}) with  $\rho=20$ .}
\label{nettc:alpha}
\end{figure}

\begin{figure}[b] 
\centering
\includegraphics[width=\linewidth]{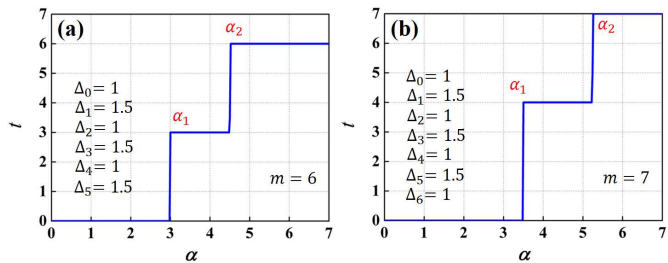}
\caption{Net topological charge $t$ of an AMRFV beam as a function of $\alpha$ with different $m$ for $\rho=20$.}
\label{tc:m}
\end{figure}

\begin{figure*}[htbp]  
\centering
\includegraphics[width=14cm]{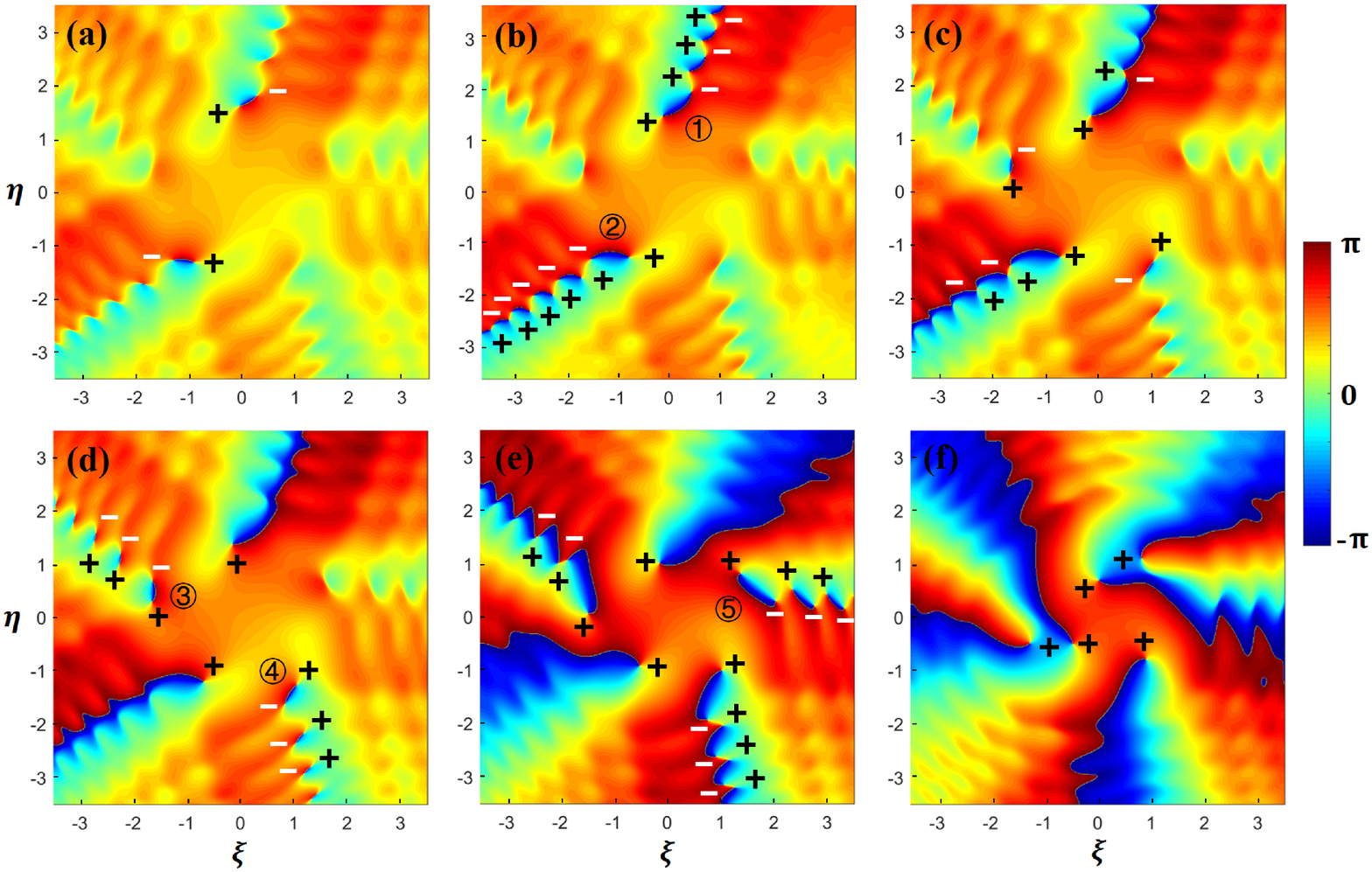}
\caption{Phase evolution of an AMRFV beam with $m=5$ and $\left[ \Delta_0, \Delta_1, \Delta_2, \Delta_3, \Delta_4 \right]=\left[ 1, 1.5, 1, 1.5, 1.75 \right] $. The symbol ``+''and ``-'' denote  unit left- and right-handed vortices, respectively. (a) $\alpha=2.1$, (b) $\alpha=2.5$, (c) $\alpha=2.75$, (d) $\alpha=2.9$, (e) $\alpha=4.3$ and (f) $\alpha=5.8$.}
\label{phaseevolution}
\end{figure*}

To illustrate the possibilities in controlling TC jump in AMRFV beams, we limit our investigation to three special cases, $m=5$, $m=6$ and $m=7$, and consider how the net topological charge depends on the other beam parameters, namely $\alpha$ and $\Delta_p$.

Figure \ref{nettc:alpha} shows how the jump in topological charge depends on the choice of the fractional phase control coefficients $\Delta_p$. From the figure, it is clear that the topological charge jumps at values of $\alpha$ such that $\alpha  = 0.5\left( {2j + 1} \right)m{\Delta _p}\left( {j = 0,1,2,...} \right)$, for every value of $p$. Obviously, the emergence of the jump has certain periodic characteristics. It is to be noted that this periodicity is not always symmetrical (simply repeating the previous cycle), and the specific periodicity is determined by the values of $\Delta_p$ and $m$. For example, Fig.~\ref{nettc:alpha}(a) is a typical repeatable cycle and  the period is 5, while Figs.~\ref{nettc:alpha}(b-d) are not. To give a clear and quantifiable description, we only focus on the situation in the first period (e.g., $0 \le \alpha  \le 5$) below, and the cases in other periods (larger $\alpha$) can be deduced by analogy. In the first period (i.e., the left side of the green line), the number of distinct jumps that occur and the size of each jump are equal to the number of different values of $\Delta_p$ and the multiplicity of $\Delta_p$, respectively. The number of different values of $\Delta_p$, for instance, in Figs.~\ref{nettc:alpha} (a)-(d) are 1 (i.e., 1), 2 (i.e., 1 and 1.5), 3 (i.e., 1, 1.5 and 1.75) and 4 (i.e., 1, 1.25, 1.5 and 1.75), respectively. Thus, the number of jumps that occur are 1, 2, 3 and 4, respectively. For Fig.~\ref{nettc:alpha}(a), there are five identical values of $\Delta_p$, so the single jump unit is 5 (which occurs at $\alpha_1=2.5$). For Fig.~\ref{nettc:alpha}(b), the multiplicities of $\Delta_p=1$ and $\Delta_p=1.5$ are 3 and 2,  respectively, thus first jump amount (at ${\alpha _{1}}=2.5$) is 3 and the second jump amount (at $\alpha_{2}=3.75$) is  2. In a similar fashion, ${\alpha _{1}}$- ${\alpha _{3}}$ in Fig.~\ref{nettc:alpha}(c) and ${\alpha _{1}}$-${\alpha _{4}}$ in Fig.~\ref{nettc:alpha}(d), as well as the jump amounts, can be determined.

Figure \ref{tc:m} shows how the topological charge jump can be adjusted by adjusting the number $m$ of multi-ramp sections. Comparing Fig.~\ref{nettc:alpha}(b) and Fig.~\ref{tc:m}, we find that although both figures have only two values of $\Delta_p$ (i.e., 1 and 1.5), the choice of different $m$ leads to different jump locations and different jump units.

From Figs. \ref{nettc:alpha}-\ref{tc:m}, it can be concluded that
\begin{equation}  
t = \sum\limits_{p = 0}^{m - 1} {{\rm{Int}}\left[ {\frac{\alpha }{{m{\Delta _p}}} + \frac{1}{2}} \right]} ,
\label{tc:math}
\end{equation}
where Int denotes the integer arithmetic.  This formula is surprisingly straightforward and makes it relatively easy to choose the number of ramps $m$ and the values $\Delta_p$ to produce any evolution of topological charge desired.

The means by which the topological charge jumps may be seen to be a ``Hilbert Hotel'' evolution of the number of singularities in the system, as first noted in \cite{gg:o:2016}. In Fig.~\ref{phaseevolution}, we calculate the phase evolution of an AMRFV beam with the same parameters as in Fig.~\ref{nettc:alpha}(c). An optical vortex is represented by a point where all the colors converge -- a singularity of phase. Left- and right-handed vortices (corresponding  to positive and negative TCs) are represented by the phase increasing or decreasing in a counterclockwise path around the singularity, respectively.

As $\alpha$ increases from zero to ${\alpha _{1}} =0.5m\Delta _0= 2.5$,  pairs of vortices with equal and opposite TCs begin to appear, with additional pairs appearing along two lines extending away from the origin [Fig.~\ref{phaseevolution}(a)]. These pairs contribute nothing to the net topological charge $t$, which remains $t=0$. For $\alpha {\rm{ = }}{\alpha _{1}} = 2.5$, two infinite lines of vortex pairs ( labeled {\Large{\textcircled{\small{1}}}} and {\Large{\textcircled{\small{2}}}}) have appeared, corresponding to the two values ${\Delta _0}{\rm{ = }}{\Delta _2} = 1$, as seen in Fig.~\ref{phaseevolution}(b).  As $\alpha$ increases past $\alpha {\rm{ = }}{\alpha _{1}} = 2.5$, the negatively-charged vortices annihilate with their more distant positive neighbor, leaving the net topological charge $t=2$ [Fig.~\ref{phaseevolution}(c)]. At the same time, two new finite vortex chains (i.e.,{\Large{\textcircled{\small{3}}}} and {\Large{\textcircled{\small{4}}}}) begin to grow due to the two values ${\Delta _1}{\rm{ = }}{\Delta _3} = 1.5$ [Fig.~\ref{phaseevolution}(d)]. As $\alpha$ increases from  $\alpha_{2}=0.5m\Delta _1=3.75$ to  $\alpha_{3}=0.5m\Delta _4=4.375$ [Fig.~\ref{phaseevolution}(e)],  vortex chains {\Large{\textcircled{\small{3}}}} and {\Large{\textcircled{\small{4}}}} contribute two vortices with positive TC and we finally have $t=4$. Again during the annihilation of vortex chains {\Large{\textcircled{\small{3}}}} and {\Large{\textcircled{\small{4}}}} , a new vortex chain {\Large{\textcircled{\small{5}}}} is created due to the ramp with $\Delta_4=1.75$. Finally , as $\alpha$ increases past $\alpha {\rm{ = }}{\alpha _{3}} = 4.375$ [Fig.~\ref{phaseevolution}(f)], all the annihilations are done and the net topological charge $t=5$. Through the above analysis we know that the evolution process in Fig.~\ref{phaseevolution} is in complete agreement with that in Fig.~\ref{nettc:alpha}(c) and Eq. (\ref{tc:math}).  

More specifically, Eq. (\ref{tc:math}) can be further explained by the evolution of vortices, namely, as the equivalent TC $\alpha /{\Delta _p}$ in each ramp section gradually approaches the values of $m/2$, a line of vortices in each ramp section is created, for instance, as can be seen in Fig.~\ref{phaseevolution}(a-b) for the evolution of the vortex chain {\Large{\textcircled{\small{1}}}} or {\Large{\textcircled{\small{2}}}}. After $\alpha /{\Delta _p}=m/2$, we are left with an unbalanced vortex with positive TC in each corresponding ramp section which leads to TC jump, for instance, as can be seen in Fig.~\ref{phaseevolution}(d) for the remaining individual vortex after the annihilation of {\Large{\textcircled{\small{1}}}} or {\Large{\textcircled{\small{2}}}}.  

It is to be noted that our results apply to an ideal plane wave, or a beam that is effectively wide enough to be approximated as a plane wave. It has been shown recently for both traditional fractional vortex sources \cite{jwlwxyjzsz:oe:2019} as well as multi-ramp fractional vortex sources \cite{jwbggzycsyzlgw:olt:2020} that for narrow width beams, the jumps in topological charge occur at integer, and not half-integer, source values. The implications for AMRV beams of finite width will be considered in separate work.

In this letter, we have shown how even an optical element with a single tunable parameter can be used to generate rich topological behavior; if one considers an AMRSPP with $m$ sections each with an independently tunable ramp charge $\alpha_p$, Eq.~(\ref{tc:math}) can be generalized to the form,

\begin{equation}  
t = \sum\limits_{p = 0}^{m - 1} {\rm{Int}}\left[ \frac{\alpha_p }{m} + \frac{1}{2} \right].
\label{tc:math2}
\end{equation}

If we further allow the azimuthal width of the $m$ sections to vary, with the constraint that the total equals $2\pi$, it seems clear that we can then also adjust the azimuthal positions of the vortex chains, which will roughly coincide with the discontinuity lines between sections.  By generalizing Eqs.~(\ref{plate:def})-(\ref{amrfv:field}) to take into account $m$ aziumuthal sections of width $2\pi/m_p$, we may further generalize Eq.~(\ref{tc:math2}) to the form,
\begin{equation}
t = \sum\limits_{p = 0}^{m - 1} {\rm{Int}}\left[ \frac{\alpha_p }{m_p} + \frac{1}{2} \right].
\end{equation}
Both the azimuthal positions of the vortex chains, as well as their evolution as a function of $\alpha_p$, will depend on the choice of $m_p$.

This letter therefore shows that it is possible to generalize the Hilbert Hotel vortex creation of Ref.~\cite{gg:o:2016} through the use of an anomalous multi-ramp spiral phase plate, allowing an arbitrary number of vortices to be created at any value of the source charge $\alpha$.  These results show that we have great flexibility in controlling the topological charge of an AMRFV beam, and can readily predict and plan the appearance of new vortices. We have revealed the relationships between the value where the jump appears, the number of jumps that occur, the unit of each jump of an AMRFV beam and beam parameters. Our work may be useful in multi-particle trapping, information encoding and optical devices manufacturing, and highlights the richness of possibilities in vortex creation.\\

\noindent{\bf Funding.} National Key Research and Development Program of China (2019YFA0705000); National Natural Science Foundation of China (NSFC) (91750201 \& 11525418 \& 11974218 \& 11774250); Innovation Group of Jinan (2018GXRC010);  China Scholarship Council (201906920047); Tang Scholar.  Greg Gbur was funded by ONR MURI N00014-20-1-2558.\\

\noindent\textbf{Disclosures.} The authors declare no conflicts of interest.

\clearpage

{\bf {FULL REFERENCES}}
\begin{enumerate}
\item G. J. Gbur, \emph{Singular Optics} (CRC, 2017).
\item L. Allen, M. W. Beijersbergen, R. J. C. Spreeuw, and J. P. Woerdman,``Orbital angular momentum of light and the transformation of Laguerre-Gaussian laser modes,''  Phys. Rev. A   \textbf{45}, 8185 (1992).
\item G. Gbur and R.K. Tyson, ``Vortex beam propagation through atmospheric turbulence and topological charge conservation,'' J. Opt. Soc. Am. A \textbf{25}, 225 (2008).
\item M. V. Berry, ``Optical vortices evolving from helicoidal integer and fractional phase steps,'' J. Opt. A: Pure Appl. Opt. \textbf{6}, 259 (2004).

\item I.V. Basistiy, M.S. Soskin and M.V. Vasnetsov, ``Optical wavefront dislocations and their properties,'' Opt. Commun. 119, 604 (1995).

\item J. Leach, E. Yao, and M. J. Padgett, ``Observation of the vortex structure of a non-integer vortex beam,'' New J. Phys. \textbf{6}, 71 (2004).

\item C. Guo, Y. Yu, and Z. Hong, ``Optical sorting using an array of optical vortices with fractional topological charge,'' Opt. Commun. \textbf{283}, 1889 (2010).
\item Y. Fang, Q. Lu, X. Wang, W. Zhang, and L. Chen, ``Fractional-topological-charge-induced vortex birth and splitting of light fields on the submicron scale,'' Phys. Rev. A \textbf{95}, 023821 (2017).
\item J. Zeng, X. Liu, F. Wang, C. Zhao, and Y. Cai,``Partially coherent fractional vortex beam,''  Opt. Express \textbf{26}, 26830 (2018).
\item S. Li, B. Shen, X. Zhang, Z. Bu, and W. Gong, ``Conservation of orbital angular momentum for high harmonic generation of fractional vortex beams,'' Opt. Express \textbf{26}, 23460 (2018).
\item J. Wen, L. Wang, X. Yang, J. Zhang and S. Zhu, ``Vortex strength and beam propagation factor of fractional vortex beams,'' Opt. Express \textbf{27}, 893 (2019).

\item J. Zeng, C. Liang, H. Wang, F. Wang, C. Zhao, G. Gbur and Y. Cai, ``Partially coherent radially polarized fractional vortex beam,'' Opt. Express \textbf{28}, 11493 (2020).
\item M. Beijersbergen, R. Coerwinkel, M. Kristensen, and J. Woerdman, ``Helical-wavefront laser beams produced with a spiral phaseplate,'' Opt. Commun. \textbf{112}, 321 (1994).
\item G. Gbur, ``Fractional vortex Hilbert's Hotel,'' Optica  \textbf{3}, 222 (2016).
\item X. Fang, H. Ren, and M. Gu, ``Orbital angular momentum holography for high-security encryption,'' Nat. Photonics \textbf{14}, 102 (2020).
\item Y. Ren, Z.Wang, P. Liao, L. Li, G. Xie, H. Huang, Z. Zhao, Y. Yan, N. Ahmed, A.Willner, M. P. J. Lavery, N. Ashrafi, S. Ashrafi, R. Bock, M. Tur, I. B. Djordjevic, M. A. Neifeld, and A. E. Willner, ``Experimental characterization of a 400 Gbit/s orbital angular momentum multiplexed free-space optical link over 120 m,'' Opt. Lett. \textbf{41}, 622 (2016).

\item S. Wang, and D. Zhao, \emph{Matrix Optics} (Springer, 2000).


\item J. Wen, B. Gao, G. Zhu, Y. Cheng, S. Zhu, L. Wang,
``Observation of multiramp fractional vortex beams and their total vortex strength in free space,'' Opt. \& Laser Tech.
\textbf{131}, 106411 (2020).
\end{enumerate}


\begin{thebibliography}{1}
\bibitem{gjg:so:2017} G. J. Gbur, \emph{Singular Optics} (CRC, 2017).
\bibitem{lamwbrjcsjpw:pra:1992} L. Allen, M. W. Beijersbergen, R. J. C. Spreeuw, and J. P. Woerdman, Phys. Rev. A  \textbf{45}, 8185 (1992).
\bibitem{ggrkt:josaa:2008} G. Gbur and R.K. Tyson, ``Vortex beam propagation through atmospheric turbulence and topological charge conservation,'' J. Opt. Soc. Am. A \textbf{25}, 225 (2008).
\bibitem{mvb:joa:2004} M. V. Berry, J. Opt. A: Pure Appl. Opt. \textbf{6}, 259 (2004).
\bibitem{ivbmssmvv:oc:1995} I.V. Basistiy, M.S. Soskin and M.V. Vasnetsov, Opt. Commun. \textbf{119}, 604 (1995).

\bibitem{jleymjp:njp:2004} J. Leach, E. Yao, and M. J. Padgett, New J. Phys. \textbf{6}, 71 (2004).


\bibitem{cgyyzh:oc:2010} C. Guo, Y. Yu, and Z. Hong, Opt. Commun. \textbf{283}, 1889 (2010).
\bibitem{yfqlxwwzlc:pra:2017} Y. Fang, Q. Lu, X. Wang, W. Zhang, and L. Chen, Phys. Rev. A \textbf{95}, 023821 (2017).
\bibitem{jzxlfwczyc:oe:2018} J. Zeng, X. Liu, F. Wang, C. Zhao, and Y. Cai, Opt. Express \textbf{26}, 26830 (2018).
\bibitem{slbsxzzbwg:oe:2018} S. Li, B. Shen, X. Zhang, Z. Bu, and W. Gong, Opt. Express \textbf{26}, 23460 (2018).
\bibitem{jwlwxyjzsz:oe:2019} J. Wen, L. Wang, X. Yang, J. Zhang and S. Zhu, Opt. Express \textbf{27}, 893 (2019).


\bibitem{jzclhwfwczggyc:oe:2020} J. Zeng, C. Liang, H. Wang, F. Wang, C. Zhao, G. Gbur and Y. Cai, Opt. Express \textbf{28}, 11493 (2020).
\bibitem{mbrcmkjw:oc:1994}M. Beijersbergen, R. Coerwinkel, M. Kristensen, and J. Woerdman, Opt. Commun. \textbf{112}, 321 (1994).
\bibitem{gg:o:2016} G. Gbur, Optica  \textbf{3}, 222 (2016).
\bibitem{xfhrmg:np:2020}X. Fang, H. Ren, and M. Gu, Nat. Photonics \textbf{14}, 102 (2020).
\bibitem{yrzwplllgxhhzzyynaawmpjlnasarbmtibdmanaew:ol:2016}Y. Ren, Z.Wang, P. Liao, L. Li, G. Xie, H. Huang, Z. Zhao, Y. Yan, N. Ahmed, A.Willner, M. P. J. Lavery, N. Ashrafi, S. Ashrafi, R. Bock, M. Tur, I. B. Djordjevic, M. A. Neifeld, and A. E. Willner, Opt. Lett. \textbf{41}, 622 (2016).


\bibitem{swdz:mo:2000}S. Wang, and D. Zhao, \emph{Matrix Optics} (Springer, 2000).


\bibitem{jwbggzycsyzlgw:olt:2020} J. Wen, B. Gao, G. Zhu, Y. Cheng, S. Zhu, L. Wang, Opt. \& Laser Tech. \textbf{131}, 106411 (2020).
\end{thebibliography}
\end{document}